# Trends in Banking 2017 and onwards

**Peter Mitic** [1,2,3]

[1] Santander UK
2 Triton Square, Regent's Place, London NW1 3AN
Email: peter.mitic@santander.co.uk
Tel: +44 (0)207 756 5256

[2] Department of Computer Science, University College London
Gower Street, London WC1E 6BT

[3] Laboratoire d'Excellence sur la Régulation Financière (LabEx ReFi)

**ABSTRACT**

*The changing nature of the relationship between a retail bank and its customers is examined, particularly with respect to new financial concepts, debt and regulation. The traditional image of a bank is portrayed as a physical building a classical Doric portico. This image conveys concepts of service, soundness, strength, stability and security ("five-S"). That "five-S" concept is changing, and the evidence for changes that affect customers directly is considered. A fundamental legal problem associated with those changes is highlighted: a bank is no longer solely responsible for the safeguard of customer monies. A solution to this problem is proposed: banks should be jointly liable with perpetrators of criminal activity in the event of frauds as an encouragement to recognise and mitigate fraud.*

# 1 INTRODUCTION AND MOTIVATION

The traditional image of a bank is clear from a Google search on 'bank', and choosing the *image* category. It's a Doric-inspired portico: a solid physical building that is designed to impress and inspire confidence. Figure 1 shows a sample. When you see that image, you think "BANK".

*Figure 1: Traditional images of a bank*

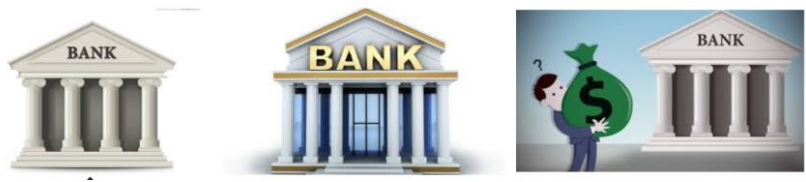

The 'bank' image is one of *Service, Soundness, Strength, Stability and Security*, which we call "*five-S*" from now on. This paper examines the changing nature of the *five-S* image. The last *S*, presents a tricky legal problem: with remote banking, a bank is no longer legally the sole custodian of customer monies. The customer is too, which puts the customer at a huge disadvantage in the event of fraud. This paper is organised in five main sections (2-6), one for each *S*. Each one has the author's projection for the *S* in question for the next few years.

# 2 SERVICE: Branches, Money, the Internet and Innovation

In this section we discuss how the services provided by a bank is likely to change.

## 2.1 Branches and Mobile banking

An increasing trend to use digital media for banking transactions instead of branch usage is said to be the main cause of the demise of bank branches over recent years. Effectively customers are forced to use the internet, whether or not they like it. This point has been investigated for the UK Parliament (Edmonds 2016), and the figures in Tables 1 and 2 are reported.

*Table 1: Numbers of branches 1988-2012*

|  | *1988* | *1995* | *2000* | *2005* | *2010* | *2012* |
|---|---|---|---|---|---|---|
| *Banks & Building Societies* | 20583 | 16723 | 11026 | 10232 | 9309 | 8837 |
| *Post Offices* | n/a | n/a | 18393 | 14609 | 11905 | 11500 |
| *Total* |  |  | 29419 | 24841 | 21214 | 20337 |

*Table 2: Bank closures January 2015 - December 2016*

| *Bank* | *Closures* | *% of network* |
|---|---|---|
| *HSBC* | 321 | 27 |
| *RBS/NatWest* | 191 | 10 |
| *Lloyds* | 180 | 14 |
| *Barclays* | 132 | 8 |
| *Co-operative* | 117 | 53 |
| *Santander* | 87 | 8 |
| *TSB* | 18 | 3 |

Bank network contraction is driven by cost cutting, mergers within the industry, competition from 'challenger banks', and growth in alternative means of accessing bank services. Closures are acute in poorer, mainly rural, areas where the economic value of transactions is low, and the impact on non-mobile customers is high. In parallel with branch closures, the British Bankers Association (BBA) has produced many headline figures to illustrate the growth of alternative means of banking (BBA 2016). Among them are:

- 40000 app downloads per day in 2015 (25% rise from 2014)
- 15 million contactless cards issued in 2015 – a rise of 54% from 2014.
- 250% annual rise in spending using contactless cards with £1.1bn spent in March 2016
- 11 million banking app logins in 2015 – a rise of 50% since 2014.
- 347 million payments done using banking apps in 2015 – a rise of 54% since 2014.

Branch closures also imply that fewer ATMs are available for cash withdrawals, although there is some offset not attached to banks. The use of non-branch banking has also resulted in a decline in dependence on paper statements. Customers are encouraged to receive statements on-line. A cautionary note comes from Wu and Saunders (2016). They cite the case of a recipient of statements by email which were 'lost' among many other emails, ending in legal action against the customer.

## 2.2 Cash and Cards

A further creeping change is that less cash is being used, as the BBA figure on contactless card usage hints at. The British Retail Consortium (BRC) reports the figures in Table 3 on cash and card usage (BRC 2015).

*Table 3: Decline in cash and debit card usage 2013-4*

|  | % of sales turnover | % of sales transactions | Cost per transaction (pence) |
|---|---|---|---|
| Cash | -0.9 | -4.5 | 1.22 |
| Debit cards | -1.9 | -0.1 | 9.46 |
| Credit & Charge cards | 6.7 | 5.2 | 33.85 |

The surprising figures in table 3 are the costs per transaction. Cash is cheapest for retailers, and costs of other methods are passed to the customer. There is usually no incentive at the point of sale for the customer to compare the price of different payment methods. Despite a higher cost, some customers still prefer to use a credit card because it is a means to defer payment. Figure 2 shows a breakdown of payment methods in 2015 (source: http://www.thisismoney.co.uk/money/saving/article-3604494/ Britons-write-HALF-BILLION- cheques-year-says-Payments-Council.html#ixzz4bccsnsNc).

*Figure 2: Payment methods*

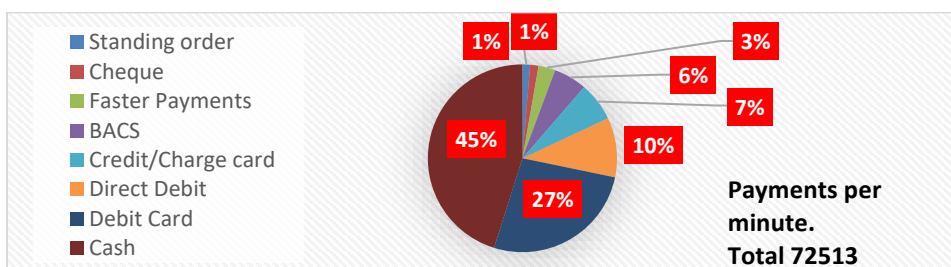

The problem for the banks is one of being sensitive to customer needs whilst remaining competitive. Banks are forced into promoting non-branch banking because that is what competitors do. Customers who prefer branch usage, cheques etc. are sometimes ignored. Curry and Penman (2004) argue for a balance between retaining personal service (i.e. branches with staff who know their customers) and electronic (i.e. machine interaction) banking as a means to attract and retain customers.

## 2.3 Innovations

Advances have been made in automated decision-making methodologies, and there are some projections that many manual tasks will be machine-controlled after about five years from now. Innovations concentrate on 'customer experience', impacting in the following areas.

- Assessment of special services: approval of new accounts, loans and mortgages. The purely deterministic processes that result in "Computer says NO!" are not acceptable.
- Advancement in the sophistication of so-called 'robo-advisors' for investment advice, mainly using pattern-matching techniques, and superseding the analyst's ability to define use-cases.
- Adaptive systems that can "learn" from new data.
- Fraud detection by identifying unusual transactions, patterns and styles.

## 2.4 Reputation

The underlying assumption behind the reputation management is that it is a worthwhile activity. This may seem obvious, but it has never been demonstrated by quantitative means until now (Mitic 2017b). The indications are that negative reputational events have a much more serious impact than positive reputation of events. Increasing interest in the measurement of reputation indicates that reputation will be of particular interest in the years 2017-18. The principal driver for improving reputation is that reputation affects sales, profits, new business, conduct (which translates into regulatory fines) and bond issues. Furthermore, the reputation of any one bank affects all banks in the sector (Mitic 2017a).

## 2.5 Projections: Service

- Continued branch closures and increased reliance on internet banking and mobile devices.
- Long term, the emergence of a few multi-bank branches in major areas of population, with none in rural areas.
- Increasing use of card (including contactless) payment
- Long term, phasing out of cash
- Adaptive systems that can "learn" from new data still do not mimic the function of the human brain in decision making.
- Behind-the-scenes reputation management (not visible to customers), not quantified.

## 3 SOUNDNESS: Regulation and Conduct

This sections deals with regulation in the context of the current economic climate.

## 3.1 Regulatory Fines

There has been a pattern of reaction in the form of regulation to successive events ever since the inception of regulation in 1977 - EC directive 77/780 (http://eur-lex.europa.eu/legal-content/EN/TXT/?uri=CELEX%3A31977L0780). The most recent guidance is in 2013 (CII 2013).

Each new banking crisis has been followed shortly after by new banking legislation or a reorganisation of the regulatory environment. Some discussion may be found in MacConachie (2009). Financial institutions do not always stick to regulations, and fines continue to be levied. Table 4 shows the number and sum of fines annually from 2002-16. The *Source* column references the FSA fines tables (see http://www.fsa.gov.uk/about/press/facts/fines/2002) and the FCA fines tables (see https://www.fca.org.uk/news/news-stories/<<yyyy>>-fines (where <<yyyy>> = 2013 to 2017).

*Table 4: Regulatory fines 2002-16*

| Year | Number of fines | Sum of fines (£m) | Mean fine (£m) | Source |
| --- | --- | --- | --- | --- |
| 2002 | 8 | 7.44 | 0.83 | FSA |
| 2003 | 17 | 10.98 | 0.65 | FSA |
| 2004 | 32 | 24.77 | 0.77 | FSA |
| 2005 | 80 | 17.74 | 0.85 | FSA |
| 2006 | 27 | 13.31 | 0.49 | FSA |
| 2007 | 23 | 5.34 | 0.23 | FSA |
| 2008 | 51 | 22.71 | 0.45 | FSA |
| 2009 | 42 | 35.01 | 0.83 | FSA |
| 2010 | 80 | 88.52 | 1.12 | FSA |
| 2011 | 60 | 66.14 | 1.12 | FSA |
| 2012 | 53 | 311.57 | 5.88 | FSA |
| 2013 | 54 | 476.1 | 8.82 | FSA and FCA |
| 2014 | 40 | 1471.43 | 36.79 | FCA |
| 2015 | 40 | 905.22 | 22.63 | FCA |
| 2016 | 23 | 22.22 | 0.97 | FCA |
| Totals | 630 | 3478.5 | 5.52 | |

The peak in 2014-15 is due mainly to LIBOR-related fines. The largest fine recorded to date is $16.7bn, against Bank of America, for misleading investors into buying toxic mortgage securities (http://www.telegraph.co.uk/finance/newsbysector/banksandfinance/ 11047586/Bank-of-America-pays-record-17bn-to-settle-US-toxic-mortgage-claims.html). Such huge fines indicate that the regulatory environment, and fines in particular, does not deter.

### 3.2 Regulation: looking ahead

The future of regulation is uncertain in changing political and economic environments. There is increased emphasis on stress testing to ensure that adequate capital buffers are held, but this has a direct effect on banks' ability to lend. In 2017 the ECB is set to initiate new legislation on capital requirements, bank recovery and resolution, especially with a view to recapitalise Italian banks. There may be moves for greater union in European capital markets.

### 3.3 Projections: Soundness

- Static regulatory environment with increased emphasis on stress testing.
- Continuing regulatory breaches as we ponder the next significant banking scandal!

## 4   STRENGTH: Competition

Competition has been a major factor in recent years as a means of forcing banks to cut costs and offer products which they hope will give them an edge in the market. All too often, the term 'cutting costs' implies job losses, which amounts to a transference of cost elsewhere. Competition comes from two main sources: 'challenger' banks and new technology that enables peer-to-peer transactions.

### 4.1   Challenger banks

The concept of a 'challenger' bank (Shawbrook, Metro etc.) has gained increasing momentum in the past 5 years. In addition, organisations normally engaged in other activities have entered the market: supermarkets such as Tesco and Sainsbury have well-established credit card and loan facilities.

Challenger banks typically try to offer a service (personal contact, innovative products, good interest rates etc.) that marks them as distinct from the well-established high street banks. Traditional banks try to copy them, both in terms of products and innovation. Many challenger banks can compete because they do not have the heavy overheads (mainly people and premises) of the traditional banks. Supermarkets are in between traditional and challenger banks: they may be offering banking services to draw in customers. Offering low savings rates continues to be a problem for all banks since they are not attractive for savers. Table 5 shows interest rates (current in March 2017) on 1-year bonds or savings accounts offered by a selection of banks. All are low compared to the prevalent 10% rates in the 1980s and 1990s (BoE 2017). Less tangible factors, such as a convenient location (which could be the internet), an app that "works", or a reputation for quality of service can also be a significant factor in distinguishing one bank from another.

*Table 5: 1-year bond/1-year savings interest rates, March 2017*

| Bank | AER rate (%) | Type |
|---|---|---|
| Barclays | 0.55 | Traditional |
| HSBC | 0.6 | Traditional |
| Nationwide | 0.65 | Traditional |
| RBS | 0.8 | Traditional |
| Tesco | 0.81 | Supermarket |
| Santander | 0.9 | Traditional |
| Skipton | 0.9 | Traditional |
| Metro | 0.9 | Challenger |
| Virgin | 1.05 | Challenger |
| Sainsbury | 1.1 | Supermarket |
| Paragon | 1.35 | Challenger |
| Zenith | 1.36 | Challenger |
| Shawbrook | 1.45 | Challenger |
| Aldermore | 1.5 | Challenger |
| Atom | 1.6 | Challenger |

Mortgage lending has some of the same characteristics, but with the major difference that the terms and conditions associated with negotiating a mortgage are much more stringent than they were twenty years ago. Interest rate is certainly a significant factor for borrowers, but so is their credit rating, employment status, and general lifestyle. To some extent, many banks can pick who they lend to, and they prefer customers with a sound credit record and who are employed (rather than self-employed).

### 4.2 Blockchain

Blockchain technologies were originally designed to avoid banks and regulation, so it is curious that banks are starting to use them. An internet search on the keyword *blockchain* results in thousands of articles, most of which hail it as a methodology that will revolutionise banking. Most explanations of the way in which *blockchain* works delve into jargon almost immediately, and most are linked to *bitcoin*, a novel form of currency used by a *blockchain*. The aim of this section is to present a non-technical explanation of how *blockchain* without linking it to *bitcoin*, and to point out some advantages and disadvantages. It should be noted that *blockchain* is currently a "hot topic", which is largely why a discussion is included in this paper.

The *blockchain* foundation is the sentiment that banks cannot be trusted, and nor can anybody else. Blockchain is a shared public database for recording transactions that does not allow records to be altered at a later date. There is a somewhat elaborate mechanism to do it. The article Deloitte (2017) attempts to explain how it works in 100 words, and links some technical terms to everyday language. Below is our version, which avoids technicalities altogether.

I want to send you €100 using a secure electronic transfer system, and convert euros to a currency called WebMoney (WM). You don't trust me: have I really got €100? All our friends verify my balance, if they want. They all know about all my transactions. It's complicated, and the first to succeed gets paid. That person tells everybody else, and anybody who wants checks the result - for free - it's easy! If a majority is satisfied, you get your money and can convert WM to €. Things cannot be changed afterwards. Payment details are distributed to everybody.

*Bitcoin* (and similar) is a virtual ("crypto") currency, the total supply of which is determined by a mathematical algorithm rather than a sovereign state. It can be exchanged with other currencies, but is highly unstable and is therefore unreliable as a means of payment. There is some argument as to whether or not it is a genuine currency: see Guadamuz and Marsden (2015). *Blockchain* proponents stress the following advantages.

- *Trust* – there is no 'trusted intermediary' (the bank)
- *Security* – communication uses public key cryptography, currently the most secure method, and is resilient because records cannot be changed afterwards (they are *immutable*).
- The blockchain database is *distributed* to all users (additional security)
- Transactions are visible to all, which implies *interoperability* and *transparency*.
- There is a full *audit trail* because each transaction has a fixed link to a previous transaction.

There are, however, considerable problems associated with *blockchain*, as below.
- It is inefficient in two ways. First, verifying blockchain transactions requires considerable energy because the verification process is a competition in which many participate. Aste (2016) estimates that a single verification needs about 1GW per second, giving a cost of approximately $5 per transaction. Second, multiple parallel attempts to verify is inherently inefficient: only one is strictly necessary.
- A distributed database is impractical if the database is large: it is not scalable. The current size of the Bitcoin database exceeds 100MB (Bitcoin 2017).
- There is an FX risk when buying and selling a crypto-currency. FX rates can vary wildly.
- Transactions are not always verified within a nominal 10 minute window. See, for example, https://news.bitcoin.com/bitcoin-transactions-stuck/. Transactions can be 'stuck' days, and some are never verified as there is no incentive for other users to do verifications.

- *Blockchain* is not secure. The Observer (http://observer.com/2015/02/the-race-to-replace-bitcoin/) has a long article that reports significant hacks totalling about $450m.
- *Blockchain* is unregulated: a computer algorithm controls the money supply, not a central bank. The Financial Times (https://www.ft.com/content/a0a4f42e-a4b1-11e5-a91e-162b86790c58) questions the desirability of this.
- There are indications that *blockchain*/*bitcoin* may be illegal. If *bitcoin* is regarded as a currency rather than a commodity, it *bitcoin* violates Title 31 of the US Federal Code (the US$ is the only legal US currency). See Guadamuz and Marsden (2015) for a discussion. Furthermore, Article 17 of the General Data Protection Regulation (GDPR) (Regulation (EU) 2016/679), gives EU individuals the *right to erasure*: they can request erasure of personal data (http://eur-lex.europa.eu/legal-content/EN/TXT/?uri=CELEX:32016R0679). The immutability of data in a blockchain conflicts with this regulation.
- If you lose your password access to the *blockchain*, it is impossible to retrieve it, and money is lost. Nobody but you is responsible for the safekeeping of your password.
- Money laundering has been linked with *blockchain* and other similar system. Some cases are detailed in Fraud Magazine (http://www.fraud-magazine.com/article.aspx?id=4294993747). This is not surprising as there are no checks on how money enters or exits a *blockchain*.

## 4.3 Peer-to-Peer transactions

Peer-to-Peer (PtP) transactions usually are loan contracts between two individuals, without the electronic 'baggage' of *blockchain*. See http://www.moneysavingexpert.com/savings/peer-to-peer-lending for details. The lender generally gets a higher (but riskier) return than from a traditional bank savings account, and the borrower has to pay a higher interest rate than had they borrowed from a bank. All PtP lending is regulated in the UK by the Financial Conduct Authority.

The Economist (http://www.economist.com/news/special-report/21650289-will-financial-democracy-work-downturn-people-people) reports increasing growth in PtP lending since 2009. However, if interest rates rise, credit cards will probably become more competitive and PtP lending may stagnate.

## 4.4 Projections: Strength

- Crypto-currencies: too unstable for normal use, unless an exchange rate is fixed by a central organisation (e.g. the Bank of England).
- *Blockchain*: continued interest but limited usage due to lack of scalability, inefficiency and lack of any guarantee of transaction verification.
- Challenger banks: increased competition leading to tighter margins. Little change

## 5 STABILITY: Views from the Central banks

Central Banks and other organisation present periodic reports on the financial stability of the banks within their remit. In this section we consider the latest reports from the International Monetary Fund (IMF), the European Central Bank (ECB) and the Bank of England (BoE). In the United States a report comes from the Office of Financial Research (OFR), which advises Congress and the public.

## 5.1 The International Monetary Fund *World* view

The most recent IMF report is from October 2016 (IMF 2016). Short term risks have diminished, despite initial concerns over a slowdown in China. Commodity prices have risen and the initial shock of Brexit has stabilised.

But medium term risks have risen. Global growth is slower and an extended period of low inflation and low interest is expected. The political climate is unsettled in many countries where there is low income growth income and a rise in populist, inward-looking policies. As a result, legacy problems will be harder to tackle, the economy will be more exposed to shocks, and there may be a period of economic and financial stagnation. In China there is continued rapid credit growth and expanding shadow banking products are threatening financial stability.

The IMF considers that some monetary policies (e.g. negative interest rates) are reaching the limits of their effectiveness, and that there is an urgent need to implement fiscal and structural policies to raise global growth.

### 5.2 The European Central Bank *Eurozone* view

The most recent ECB report was issued in November 2016 (ECB 2016). This report points to specific problems facing the Eurozone in 2017.

Systemic stress has remained relatively low over the past six months, despite some market turbulence (e.g. Brexit and the US elections). Eurozone banks have been resilient to recent market stress, but they are still vulnerable. Banks have continued to become more risk averse, although investment funds have done the opposite. The ECB report indicates four upcoming risks for 2017-18.

1. *Global risk repricing, financial contagion and political uncertainty*. This is due to rising asset prices and asset volatility (stocks and bonds), partly fuelled by political uncertainty (e.g. Brexit).

2. *Weak bank profitability, low growth non-performing loans*. Low interest rates have damaged bank profits. As a result their stock prices have also suffered. These factors in turn have resulted in a slight widening of the negative gap between banks' return on equity and cost of equity.

3. *Increasing sovereign and private sector debt*. The projected aggregate euro area government debt-to-GDP ratio is high - close to 90% of GDP for 2017. The ECB therefore calls for continued structural and fiscal reforms in some Eurozone countries.

4. *Prospective stress in the investment fund sector*. Funds exposed to liquidity mismatches and funds operating with high leverage are particularly at risk. Alternative investment funds operating without regulatory leverage limits have the potential to contribute to systemic stress because they are a significant part (39%) of the investment sector.

### 5.3 The Bank of England *UK* view

The most recent BoE report was issued in November 2016 (BoE 2016), and makes five main points.

1. 2016 has been a challenging period, but substantial moves in financial market prices have not been amplified by the UK financial system. Sterling is down by about 12%, commercial property prices have fallen, government bond yields have fallen and inflation is increasing.
2. The outlook following the Brexit referendum outlook remains uncertain, and it will take time to clarify the UK's relationships with the EU and other countries.

3. There are increased vulnerabilities due the global environment and financial markets. In particular, there are concerns over US fiscal policy, increasing reliance on credit in China, sovereign debt in the Eurozone (particularly in Italy, where impaired loans amount to €360bn compared to €225bn in equity), and elections in some European countries.
4. The UK banking system has sufficient capital to sustain the provision of financial services. Stress tests indicate that most UK banks are resilient to interest rate changes, property market revaluations, and to severe stress in the euro area and in China.
5. Liquidity in some markets is fragile, and Brexit is an exacerbating factor.

In March 2017 the Bank of England announced details of their 2017 "worst case scenario" stress test (BoE 2017a). This scenarios paints a gloomy picture of the banking sector in 2017-18:

- UK bank rate peak at 4% in 2017
- Property price fall 33% (residential), 40% (commercial)
- GBPUSD down 27% from Q4 2106 level: to $0.85=£1
- Oil price fall to $24/barrel
- UK GDP falls to 1.2%, Chinese to 3.5%: stagnation in global trade
- UK unemployment high at 9.5%

These points reinforce current worries over global trade, Brexit and consumer spending. The "worst case" may not happen, but UK banks are being asked to prepare for it.

## 5.4 The Federal Reserve Board *US* view

The Office of Financial Research (OFR 2016) advises the FRB, and assesses U.S. financial stability risk to be medium. They highlight four themes, all flagged as a contribution to contagion risk.

1. *The effect on U.S. financial stability of disruptions in the global economy*. The main disruptions are perceived as coming from Europe, Brexit being one of them.

2. *Risk-taking amid low long-term interest rates*. Low interest rates are encouraging nonfinancial corporates to take on more debt, with the potential of turning into bad debt.

3. *Risks facing U.S. financial institutions*. Cyber-crime is prominent.

4. *Challenges to improving financial data*. Deficiencies in data collection, processing and management need to be addressed.

## 5.5 Projections: Stability

- Low interest rates, low wages, low profitability, low growth.
- Measures are in place to demonstrate the resilience of the banking system.
- Political and economic worries to continue, particularly in the UK.

## 6 SECURITY: Passwords, biometrics, fraud and the law

Security breaches, some of them widely reported in the national press, have dented the reputation of several banks. The report from Techworld (http://www.techworld.com/security/uks-most-infamous-

data-breaches-3604586/) lists the most recent significant ones. One of them, Tesco Bank in November 2016 was specifically mentioned in the security report from the Bank of England (BoE 2016). £2.5m was stolen from about from 9,000 customer accounts. The UK National Crime Agency reports (NCA 2016) that cyber-enabled fraud constituted approximately 36% of all UK crime in 2015.

## 6.1 Security innovations

Security breaches dent the reputation of the organisations concerned, and expose them to regulatory fines for misconduct. There is a significant link between reputation and conduct risk (see Mitic 2016b). Some efforts have been made to improve security measures, notably by using biometrics such as finger prints or voice recognition. Examples are Barclays Bank, First Direct and HSBC (voice recognition as a form of secure ID for telephone banking), Atom Bank (log on using face recognition), and HSBC (finger print recognition). The BBC report http://www.bbc.co.uk/news/business-36939709 has details. This BBC article contains the following quote from James Daily (founder of the consumer website *Fairer Finance*) as an answer to questions on the capabilities of voice recognition:

"*In reality, consumers should have little to fear, as banks are still liable for any fraud unless they can prove that a customer was negligent. So if this technology does lead to any increase in fraud, it will be the banks that have to pick up the bill, and not customers*."

This quote is intended to reassure customers that new security procedures are to their advantage and should be welcomed. The message is clear: biometric methodologies are quick, accurate and there is no need to remember or write down passwords and PINs (writing down passwords and PINs is likely to breach terms and conditions of holding the bank account). However, the part of the quote "*unless they can prove that a customer was negligent*" exposes a fundamental legal problem that concerns the relationship between a bank and its customers (see section 6.3).

One element is missing from the discussion above. The customer needs to be assured that the counterparty that they communicate with is, indeed, the bank, and not a fraudster. The bank needs a biometric too.

## 6.2 The change in the type of bank crime

The changing nature of banking, specifically the shift from branch-based banking to online banking, has quietly changed criminal activity with respect to banking. 'Traditional' physical bank robbery has declined, as Figure 3 shows (source https://www.bba.org.uk/news/press-releases/the-decline-of-the-british-bank-robber/#.VX29r0YXd8U).
The decline in bank robbery is attributed to improvements in branch security (more CCTV, time delays and protective screens), less cash held in branches, closer co-operation between banks and the police, and better staff training.

In contrast, incidence of cyber-crime has increased markedly. The most recent figure comes from Financial Fraud Action UK (https://www.financialfraudaction.org.uk/news/2017/03/16/financial-fraud-costing-uk-2-million-a-day/). They report that in 2016, the UK lost £2 million each day as a result of financial fraud. The overall amount lost in the UK in 2016 to financial fraud was £768.8m, an increase of 1.8% year-on-year from 2015. It is becoming harder to calculate a reliable figure for financial fraud as much is thought to go unrecorded.

*Figure 3: Decline of the number of physical bank robberies*

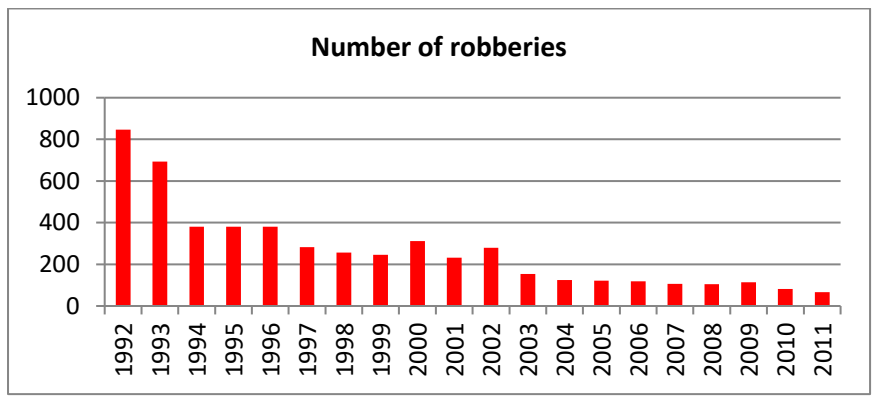

## 6.3 Security and the law

There is an important difference between cyber-crime and 'traditional' bank robbery. With cyber-crime, the *customer* is defrauded, not the bank. Individual bank accounts suffer, and the loss is not shared among all bank customers. Banks then can escape the responsibility of being the sole agent charged with the safe keeping of customer monies. With 'remote' banking the customer is also responsible, and that principle is enshrined in contract law.

The relationship between a bank and an account holder is governed by the contract (often expressed as 'terms and conditions') between them. Any breach of those terms and conditions enables a bank to avoid compensating a customer in the event of a cyber fraud. Typically, a breach might be writing down or telling somebody else a PIN or password.

In cases of cyber-crime where the customer very clearly has no involvement, such as hacking of a bank's central data store, the customer is protected by the criminal law on theft and fraud (the Theft Act 1968 and the Fraud Act 2006). Theft is the unauthorised taking of property from another, with the intent to permanently deprive that person of the property, and carries a maximum penalty of 7 years imprisonment. Fraud is the abuse of position, or false representation, or prejudicing someone's rights for personal gain, and carries a maximum penalty of 10 years imprisonment. Neither is as serious a crime as is robbery, which is associated with a degree of coercion (physical force or fear). Robbery has a maximum penalty of life imprisonment. Cyber-crime is not robbery, so it is less serious than 'traditional' bank robbery. The police treat it as such.

Cases involving frauds against individual are different. The customer is protected by the Consumer Rights Act 2015 (http://www.legislation.gov.uk/ukpga/2015/15/ pdfs/ukpga_20150015_en.pdf), but the case may have to be argued in court. Individuals rarely win court cases against a bank unless they have above average legal support.

A further complication occurred in 2014: the case of Crestsign versus Royal Bank of Scotland and NatWest in the High Court (Crestsign 2014). The case concerned termination of an interest rate swap agreement which was proving to be too expensive for Crestsign, a small business, when interest rates fell. The bank did not advise them that they would be liable if interest rates fall, and used a 'basis clause' in its contract terms and conditions to preclude any advisory duty with respect to potential termination. The judgement was in favour of the bank. The implication for the customer is profound.

A bank does not owe a *duty of care* to its customers, in contrast to professional organisations such as architects, doctors and solicitors. It need not advise on matters that the customer does not ask about. Pending an appeal, the Crestsign case was settled out of court in February 2016 (Warwick 2016). The case law is therefore unchanged, and banks can still avoid responsibility for any advice given. Further discussion may be found in Mitic (2016a).

The situation is even worse for customers who fall foul of the *vishing* fraud. This is when they are persuaded by a fraudster, pretending to be a bank employee, to transfer monies to an account which later is shown to belong to the fraudster. Banks often take the view that since the customer authorised the transfer, the customer must bear the loss. Customers can lose their life savings in this way, and banks can be very unsympathetic.

### 6.3.1 A proposal for additional legal protection for defrauded customers

A one-line summary of the previous section is that many customer who are defrauded have little legal protection. We therefore propose the following change to the law.

Both the bank and the perpetrator of a cyber-fraud should be held jointly liable

Given that it may be very difficult to find, prosecute, and convict a cyber-criminal, the bank would often be held liable. This would be an additional cost to the bank, but not an excessive one. The point is to not disadvantage the customer. A similar measure already exists for credit card purchases, for which the credit card company is liable for faulty goods or services, even though the seller of those goods or services is the origin of the fault. This proposal is also intended as an incentive to 'encourage' banks to improve security procedures so as to make fraud much harder.

### 6.4 Projections: Security

- Increased use security measures (mainly biometrics) to replace passwords, leading to the end of password usage in a few years.
- No major increase in cyber-crime with the introduction of more stringent security measures.
- Change in the law: legal protection against fraud: unlikely to happen – Parliament is too preoccupied with other matters.

## 7 DISCUSSION

An overall assessment of the "*five-S*" categories leads to a rather mixed answer to the question "Is the banking system working in 2017 or not?" Overall, the answer is "yes", as the banking sector and is not currently suffering from any problems that it cannot handle. This provides encouragement for 2017 and 2018.

The view from regulatory authorities (*Stability* and *Soundness*) is that there is sufficient resilience within the system to withstand future shocks, but regulation is not effective in preventing bad and illegal practice. Meanwhile, we are waiting to see what the next big banking scandal will be. Default on interest only loans is a possibility. Many of them will be maturing in the next few years with insufficient provision, or no provision at all, for paying back capital. Questions will be asked as to why the banks did not ensure that measures to pay back capital were in place some years ago.

There is a mixed picture for changes to levels of *Service*. Users of remote banking have seen advances that are greatly to their advantage, but others have seen traditional values eroded. The depersonalisation of banking through increased use of remote access and automation has had the further effect that KYC ("Know Your Client") procedure have to be applied more and more frequently. When automated procedures fail, no viable substitute for human content has yet presented itself. Current levels of telephone assistance are, anecdotally, inadequate. The outlook for 2017-18 looks set to be very similar to the way it has been in 2016. Significant advances in technology are still at the experimental stage.

*Security* is a parallel thread to *Service*. Changes to services offered have forced changes to security. Some of those changes are invisible to customers (such as data encryption), and are likely to improve gradually. The visible security features (logon and passwords) are set to change markedly in 2017-18 with the introduction of biometric identification. A change in the law to make the bank and the fraudster jointly liable for fraud (section 6.3.1) is probably a distant aspiration.

Competition is unlikely to contribute to *Strength* in the banking sector unless competitors can find particular markets or products that make them viable. Very few truly original products are emerging from *fintechs*, so finding niche markets or products will be difficult.